\documentclass[twoside,twocolumn,9pt]{article}
\usepackage{extsizes}
\usepackage[super,sort&compress,comma]{natbib} 
\usepackage[version=3]{mhchem}
\usepackage[left=1.5cm, right=1.5cm, top=1.785cm, bottom=2.0cm]{geometry}
\usepackage{balance}
\usepackage{times,mathptmx}
\usepackage{sectsty}
\usepackage{graphicx} 
\usepackage{lastpage}
\usepackage[format=plain,justification=justified,singlelinecheck=false,font={stretch=1.125,small,sf},labelfont=bf,labelsep=space]{caption}
\usepackage{float}
\usepackage{fancyhdr}
\usepackage{fnpos}
\usepackage[english]{babel}
\addto{\captionsenglish}{%
  
}
\usepackage{array}
\usepackage{droidsans}
\usepackage{charter}
\usepackage[T1]{fontenc}
\usepackage[usenames,dvipsnames]{xcolor}
\usepackage{setspace}
\usepackage[compact]{titlesec}
\usepackage{hyperref}
\usepackage{epstopdf}%This line makes .eps figures into .pdf - please comment out if not required.

\usepackage{amssymb}
\usepackage{gensymb}

\newcommand*{\citen}[1]{%
  \begingroup
    \romannumeral-`\x % remove space at the beginning of \setcitestyle
    \setcitestyle{numbers}%
    \cite{#1}%
  \endgroup   
}

\definecolor{cream}{RGB}{222,217,201}

\begin{document}

\pagestyle{fancy}
\thispagestyle{plain}
%\fancypagestyle{plain}{
%
%%%%HEADER%%%
%\fancyhead[C]{\includegraphics[width=18.5cm]{head_foot/header_bar}}
%\fancyhead[L]{\hspace{0cm}\vspace{1.5cm}\includegraphics[height=30pt]{head_foot/journal_name}}
%\fancyhead[R]{\hspace{0cm}\vspace{1.7cm}\includegraphics[height=55pt]{head_foot/RSC_LOGO_CMYK}}
%\renewcommand{\headrulewidth}{0pt}
%}
%%%END OF HEADER%%%

%%%PAGE SETUP - Please do not change any commands within this section%%%
\makeFNbottom
\makeatletter
\renewcommand\LARGE{\@setfontsize\LARGE{15pt}{17}}
\renewcommand\Large{\@setfontsize\Large{12pt}{14}}
\renewcommand\large{\@setfontsize\large{10pt}{12}}
\renewcommand\footnotesize{\@setfontsize\footnotesize{7pt}{10}}
\makeatother

\renewcommand{\thefootnote}{\fnsymbol{footnote}}
\renewcommand\footnoterule{\vspace*{1pt}% 
\color{cream}\hrule width 3.5in height 0.4pt \color{black}\vspace*{5pt}} 
\setcounter{secnumdepth}{5}

\makeatletter 
\renewcommand\@biblabel[1]{#1}            
\renewcommand\@makefntext[1]% 
{\noindent\makebox[0pt][r]{\@thefnmark\,}#1}
\makeatother 
\renewcommand{\figurename}{\small{Fig.}~}
\sectionfont{\sffamily\Large}
\subsectionfont{\normalsize}
\subsubsectionfont{\bf}
\setstretch{1.125} %In particular, please do not alter this line.
\setlength{\skip\footins}{0.8cm}
\setlength{\footnotesep}{0.25cm}
\setlength{\jot}{10pt}
\titlespacing*{\section}{0pt}{4pt}{4pt}
\titlespacing*{\subsection}{0pt}{15pt}{1pt}
%%%END OF PAGE SETUP%%%

%%%%FOOTER%%%
%\fancyfoot{}
%\fancyfoot[LO,RE]{\vspace{-7.1pt}\includegraphics[height=9pt]{head_foot/LF}}
%\fancyfoot[CO]{\vspace{-7.1pt}\hspace{13.2cm}\includegraphics{head_foot/RF}}
%\fancyfoot[CE]{\vspace{-7.2pt}\hspace{-14.2cm}\includegraphics{head_foot/RF}}
%\fancyfoot[RO]{\footnotesize{\sffamily{1--\pageref{LastPage} ~\textbar  \hspace{2pt}\thepage}}}
%\fancyfoot[LE]{\footnotesize{\sffamily{\thepage~\textbar\hspace{3.45cm} 1--\pageref{LastPage}}}}
%\fancyhead{}
%\renewcommand{\headrulewidth}{0pt} 
%\renewcommand{\footrulewidth}{0pt}
%\setlength{\arrayrulewidth}{1pt}
%\setlength{\columnsep}{6.5mm}
%\setlength\bibsep{1pt}
%%%%END OF FOOTER%%%

%%%FIGURE SETUP - please do not change any commands within this section%%%
\makeatletter 
\newlength{\figrulesep} 
\setlength{\figrulesep}{0.5\textfloatsep} 

\newcommand{\topfigrule}{\vspace*{-1pt}% 
\noindent{\color{cream}\rule[-\figrulesep]{\columnwidth}{1.5pt}} }

\newcommand{\botfigrule}{\vspace*{-2pt}% 
\noindent{\color{cream}\rule[\figrulesep]{\columnwidth}{1.5pt}} }

\newcommand{\dblfigrule}{\vspace*{-1pt}% 
\noindent{\color{cream}\rule[-\figrulesep]{\textwidth}{1.5pt}} }

\makeatother
%%%END OF FIGURE SETUP%%%

%%%%%%%%%%%%%%%%%%%%%%%%%%%%%%%%%%%%%%%%%%
% Alter some LaTeX defaults for better treatment of figures:
    % See p.105 of "TeX Unbound" for suggested values.
    % See pp. 199-200 of Lamport's "LaTeX" book for details.
    %   General parameters, for ALL pages:
    \renewcommand{\topfraction}{0.9}	% max fraction of floats at top
    \renewcommand{\bottomfraction}{0.8}	% max fraction of floats at bottom
    %   Parameters for TEXT pages (not float pages):
    \setcounter{topnumber}{2}
    \setcounter{bottomnumber}{2}
    \setcounter{totalnumber}{4}     % 2 may work better
    \setcounter{dbltopnumber}{2}    % for 2-column pages
    \renewcommand{\dbltopfraction}{0.9}	% fit big float above 2-col. text
    \renewcommand{\textfraction}{0.07}	% allow minimal text w. figs
    %   Parameters for FLOAT pages (not text pages):
    \renewcommand{\floatpagefraction}{0.7}	% require fuller float pages
	% N.B.: floatpagefraction MUST be less than topfraction !!
    \renewcommand{\dblfloatpagefraction}{0.7}	% require fuller float pages
    
    % remember to use [htp] or [htpb] for placement
%%%%%%%%%%%%%%%%%%%%%%%%%%%%%%%%%%%%%%%%%

%%%TITLE, AUTHORS AND ABSTRACT%%%
\twocolumn[
  \begin{@twocolumnfalse}
%\vspace{3cm}
%\sffamily
%\begin{tabular}{m{4.5cm} p{13.5cm} }

%\includegraphics{head_foot/DOI} & 
\noindent\LARGE{\textbf{Membrane morphologies induced by mixtures of arc-shaped particles with opposite curvature}} \\%Article title goes here instead of the text "This is the title"
%\vspace{0.3cm} & \vspace{0.3cm} \\

% & 
 \noindent\large{Francesco Bonazzi,\textit{$^{a}$} Carol K.\ Hall,\textit{$^{b}$} and Thomas R.\ Weikl \textit{$^{a}$}} \\%Author names go here instead of "Full name", etc.

\noindent\normalsize{Biological membranes are shaped by various proteins that either generate inward or outward membrane curvature. In this article, we investigate the membrane morphologies induced by mixtures of arc-shaped particles with coarse-grained modeling and simulations. The particles bind to the membranes either with their inward, concave side or their outward, convex side and, thus, generate membrane curvature of opposite sign. We find that small fractions of convex-binding particles can stabilize three-way junctions of membrane tubules, as suggested for the protein lunapark in the endoplasmic reticulum of cells. For comparable fractions of concave-binding and convex-binding particles, we observe lines of particles of the same type, and diverse membrane morphologies with grooves and bulges induced by these particle lines. The alignment and segregation of the particles is driven by indirect, membrane-mediated interactions.
} \\
%\end{tabular}

 \end{@twocolumnfalse} \vspace{0.6cm}

  ]
%%%END OF TITLE, AUTHORS AND ABSTRACT%%%

%%%FONT SETUP - please do not change any commands within this section
\renewcommand*\rmdefault{bch}\normalfont\upshape
\rmfamily
\section*{}
\vspace{-1cm}

%%%FOOTNOTES%%%

\footnotetext{\textit{$^{a}$~Max Planck Institute of Colloids and Interfaces, Department of Theory and Bio-Systems, Am M\"uhlenberg 1, 14476 Potsdam, Germany}}
\footnotetext{\textit{$^{b}$~North Carolina State University,  Department of Chemical and Biomolecular Engineering,  Engineering Building I, 911 Partners Way, Raleigh, NC 27695-7905, USA. }}

%Please use \dag to cite the ESI in the main text of the article.
%If you article does not have ESI please remove the the \dag symbol from the title and the footnotetext below.
%\footnotetext{\dag~Electronic Supplementary Information (ESI) available: [details of any supplementary information available should be included here]. See DOI: 00.0000/00000000.}
%additional addresses can be cited as above using the lower-case letters, c, d, e... If all authors are from the same address, no letter is required

%\footnotetext{\ddag~Additional footnotes to the title and authors can be included \textit{e.g.}\ `Present address:' or `These authors contributed equally to this work' as above using the symbols: \ddag, \textsection, and \P. Please place the appropriate symbol next to the author's name and include a \texttt{\textbackslash footnotetext} entry in the the correct place in the list.}

%%%END OF FOOTNOTES%%%

%%%MAIN TEXT%%%%

%%
\section{Introduction}

The intricately curved shapes of biological membranes are induced and maintained by a variety of proteins \cite{Shibata09,Kozlov14,McMahon15,Baumgart11}. The arc-shaped BAR domain proteins, for example, induce membrane curvature by binding to membranes \cite{Takei99,Peter04,Rao11,Mim12b,Simunovic19}. Different BAR domain proteins bind to membranes either with their inward curved, concave side or with their outward bulged, convex side and, thus, impose membrane curvature of opposite sign \cite{Qualmann11,Masuda10,Simunovic19}. Spherical and tubular membrane shapes only exhibit curvature of one sign and can be induced by a single type of proteins \cite{Frost08,Hu08,Shi15,Daum16}. Three-way junctions of tubules, in contrast, contain membrane segments with curvatures of different sign \cite{Guven14,Chen15} and are induced and stabilized by several proteins \cite{Chen12,Wang19}. The ubiquitous three-way junctions of tubules in the endoplasmic reticulum (ER) are stabilized by the protein lunapark \cite{Chen15,Schwarz16,Cui19}, while the tubules of the ER are generated by reticulon and REEP proteins  \cite{Wang19,Voeltz06,Hu08}. The protein lunapark  presumably induces a membrane curvature that is opposite to the tubular curvature generated by reticulon and REEP proteins \cite{Chen15}.  
 
In this article, we investigate the membrane morphologies induced by mixtures of arc-shaped particles that can either bind with their inward curved, concave side (``concave particles") or with their outward bulged, convex side (``convex particles").  In our coarse-grained model of membrane shaping, the membrane is described as a triangulated elastic surface, and the particles as segmented arcs. In previous Monte Carlo (MC) simulations \cite{Bonazzi19}, we found that the membrane morphologies induced by concave particles are determined by the arc angle and membrane coverage of the particles. At membrane coverages that exceed about 40\%, concave particles induce membrane tubules, irrespective of their arc angle. In MC simulations with mixtures of concave and convex particles, in contrast, we observe a large variety of morphologies that depends on the relative coverage of the different types of particles. If the membrane coverage of concave particles greatly exceeds the coverage of convex particles, we either find single membrane tubules or three tubules connected by a three-way junction. The few convex particles cluster at the three-way junctions and appear to stabilize the junction, or distort the single tubules locally. For larger fractions of convex particles, we observe lines of convex particles segregated from lines of concave particles, and membrane morphologies with grooves and bulges induced by these lines. The alignment and segregation of the convex and concave particles is driven by indirect, membrane-mediated interactions \cite{Weikl18,Idema19,Phillips09,Reynwar07} because the direct particle-particle interactions are purely repulsive in our model. A similar alignment and segregation has been previously observed in simulations with mixtures of arc-shaped inclusions of opposite curvature \cite{Noguchi17}.

\section{Methods}

\begin{figure*}
 \centering
\includegraphics[width=\linewidth]{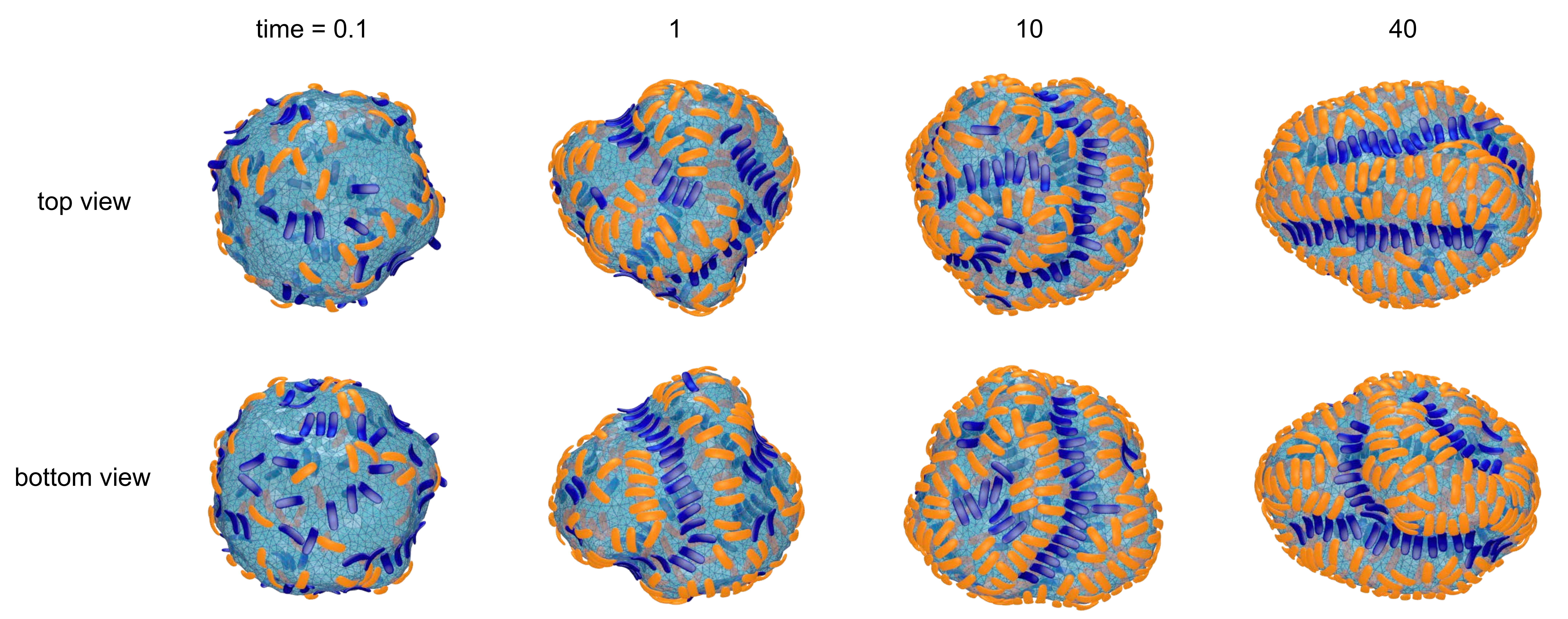}
 \caption{Time sequence of morphologies for a mixture of concave and convex particles with arc angle $60\degree$. The numbers indicate simulation times in units of $10^6$ MC steps per membrane vertex.  At time $t = 0$, the membrane has a spherical shape, and all particles are unbound. In this simulation, the adhesion energy per particle segment is $U = 11\, k_BT$, the total number of concave, orange particles is 320, and total number of convex, blue particles is 80. Only membrane-bound particles are shown in the MC snapshots. In the final morphology, 243 out of the 320 concave particles and 79 out of the 80 convex particles are bound, which leads to membrane coverages of $x_{\rm orange} = 0.37$ and  $x_{\rm blue} = 0.12$ of the particles. The reduced volume of the membrane in the final morphology is $v = 0.76$.}
 \label{figure-159}
\end{figure*}

\subsection{Model}
We model the membrane as a discretized closed surface. The bending energy of a closed continuous membrane without spontaneous curvature is the integral ${\cal E}_\text{be} = 2\kappa \oint M^2 \,dS$ over the membrane surface with local mean curvature $M$.\cite{Helfrich73}  We use the standard discretization of the bending energy for triangulated membrane described in refs.\ \citen{Julicher96,Bahrami12} and choose as typical bending rigidity the value $\kappa = 10\, k_B T$. \cite{Dimova14} Our discretized membranes are composed of $n_t = 5120$ triangles. The edge lengths of the triangles are kept within an interval $[a_m, \sqrt{3} a_m]$, and the area of the membrane is constrained to $A_0 \simeq 0.677 n_t a_m^2$ to ensure the near incompressibility of lipid membranes. \cite{Lipowsky05} The strength of the harmonic constraining potential is chosen such that the fluctuations of the membrane area are limited to less than $1\%$. The enclosed volume is unconstrained to enable a wide range of membrane morphologies with different volume-to-area ratios.

The discretized particles in our model are linear chains of 3 to 5 identical planar quadratic segments, with an angle of $30\degree$ between neighboring segments that share a quadratic edge.\cite{Bonazzi19} The arc angle of the particles, i.e.\ the angle between the first and last segment, then adopts the values $60\degree$, $90\degree$, $120\degree$ for particles composed of 3, 4, and 5 segments respectively. Each planar segment of a particle interacts with the nearest triangle of the membrane {\em via} the particle-membrane adhesion potential\cite{Bonazzi19}
\begin{equation}
V_\text{pm} = \pm U f_r(r) f_\theta(\theta)
\label{Vpm}
\end{equation}
Here, $r$ is the distance between the center of the segment and the center of the nearest triangle, $\theta$ is the angle between the normals of the particle segment and this membrane triangle, and $U>0$ is the adhesion energy per particle segment. The distance-dependent function $f_r$ is a square-well function that adopts the values $f_r(r) = 1$ for $0.25\, a_m < r <  0.75\, a_m$ and $f_r(r)=0$ otherwise. The angle-dependent function $f_\theta$ is a square-well function with values $f_\theta(\theta) = 1$ for $|\theta| < 10\degree$ and $f_\theta(\theta) = 0$ otherwise. By convention, the normals of the membrane triangles are oriented outward from the enclosed volume of the membrane, and the normals of the particle segments are oriented away from the center of the particle arc. For a negative sign in Eq.\ (\ref{Vpm}), the particles bind with their inward curved, concave surface to the membrane (``concave particles"). For a positive sign in Eq.\ (\ref{Vpm}), the particles bind with their outward bulged, convex surface to the membrane (``convex particles"). The overlapping of particles is prevented by a purely repulsive hard-core interaction that only allows distances between the centers of the planar segments of different particles that are larger than $a_p$. The hard-core area of a particle segment thus is $\pi a_p^2/4$. We choose the value $a_p = 1.5 a_m$ for the linear size of the planar particle segments. The particle segments then are slightly larger than the membrane triangles with minimum side length $a_m$, which ensures that different particle segments bind to different triangles.

\begin{figure*}
 \centering
\includegraphics[width=\linewidth]{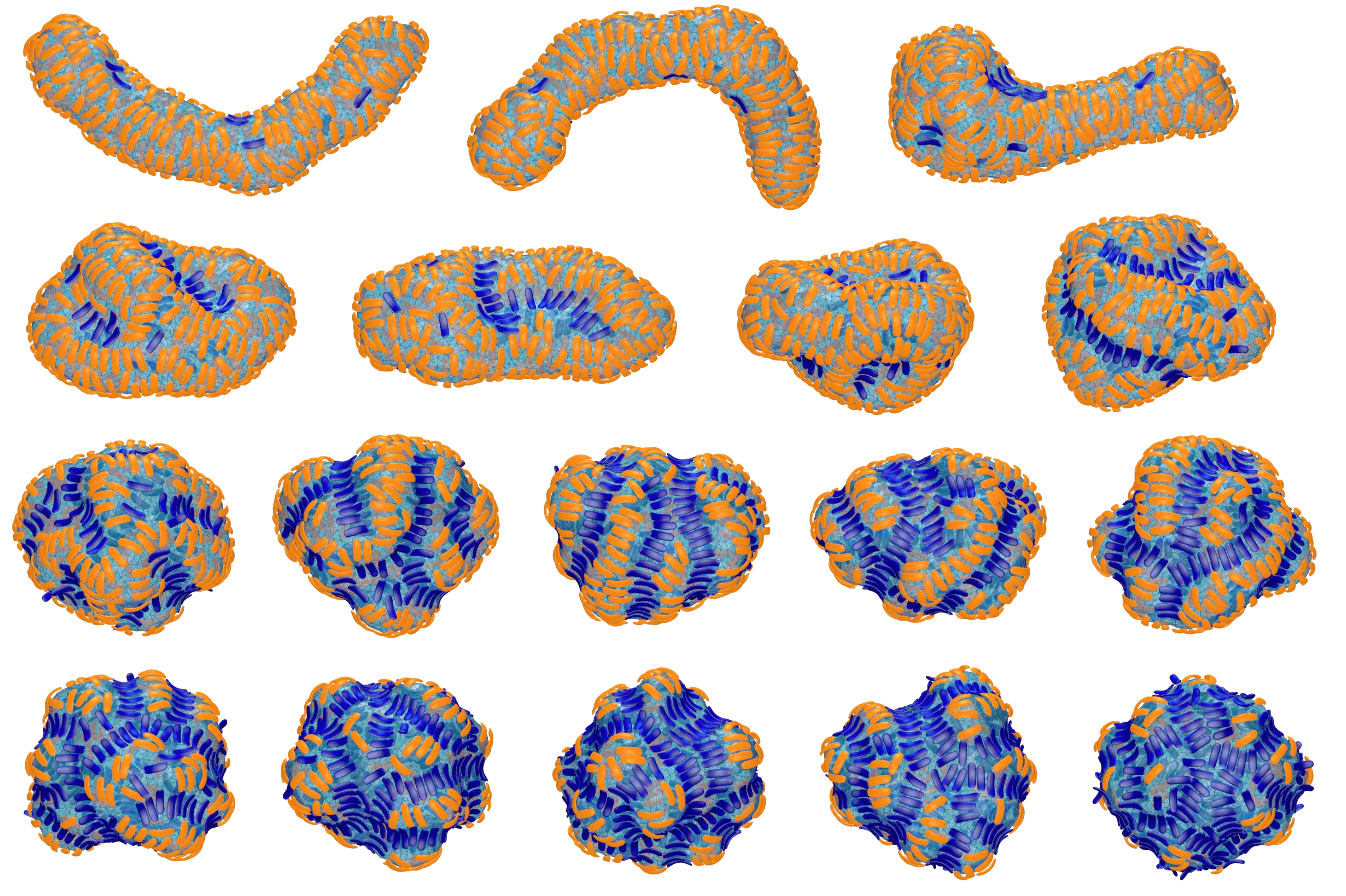}
 \caption{Representative converged morphologies for mixtures of concave and convex particles with arc angle $60\degree$. The morphologies are arranged in ascending order of the membrane coverage of convex, blue particles, which increases from top left to bottom right as $x_{\rm blue} = 0.01, 0.01, 0.03, 0.06, 0.06, 0.06, 0.12, 0.13, 0.21, 0.24, 0.24, 0.24, 0.27, 0.33, 0.36, 0.36$, and $0.37$. The membrane coverage of concave, orange particles is $x_{\rm orange} = 0.48, 0.49, 0.45, 0.43, 0.44, 0.44, 0.38, 0.30, 0.28, 0.28, 0.27, 0.27, 0.19, 0.17, 0.17, 0.16$, and $0.08$ from top left to bottom right, and the reduced volume of the membrane is $v= 0.53, 0.52, 0.63, 0.72, 0.62, 0.74, 0.75, 0.81, 0.82, 0.81, 0.81, 0.82, 0.82, 0.81, 0.81, 0.81$, and $0.83$. The morphologies result from simulations with an initially spherical membrane and the adhesion energy per particle segment $U = 10, 13, 11, 10, 13, 12, 12, 9, 10, 12, 11, 13, 10, 11, 13, 12$, and $11\, k_B T$. The overall number of bound and unbound concave particles is $392, 392, 380, 360, 360, 360, 320, 240, 240, 240, 240, 240, 160, 160, 160, 160$, and $80$ in these simulations. The total number of concave and convex particles is $400$ in all simulations.
 }
 \label{figure-3-3}
\end{figure*}
\begin{figure*}[t]
 \centering
\includegraphics[width=\linewidth]{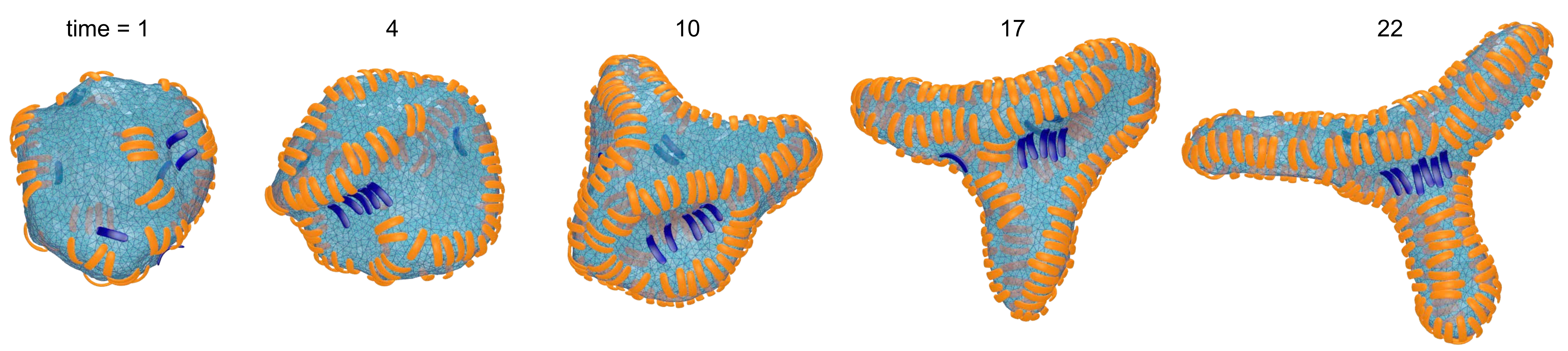}
 \caption{Time sequence of morphologies for a mixture of many concave and few convex particles with arc angle $90\degree$. The numbers indicate simulation times in units of $10^6$ MC steps per membrane vertex.  At
time $t = 0$, the membrane has spherical shape, and all particles are unbound. In this simulation, the adhesion energy per particle segment is $U = 10\, k_BT$, the total number of concave, orange particles is 392, and total number of convex, blue particles is 8. Only membrane-bound particles are shown in the MC snapshots. In the final morphology, 206 out of the 392 concave particles and all 8 convex particles are bound, which leads to membrane coverages of $x_{\rm orange} = 0.42$ and  $x_{\rm blue} = 0.016$ of the particles. The reduced volume of the membrane in the final morphology is $v = 0.49$.}
 \label{figure-209}
\end{figure*}

\begin{figure*}[t]
 \centering
\includegraphics[width=\linewidth]{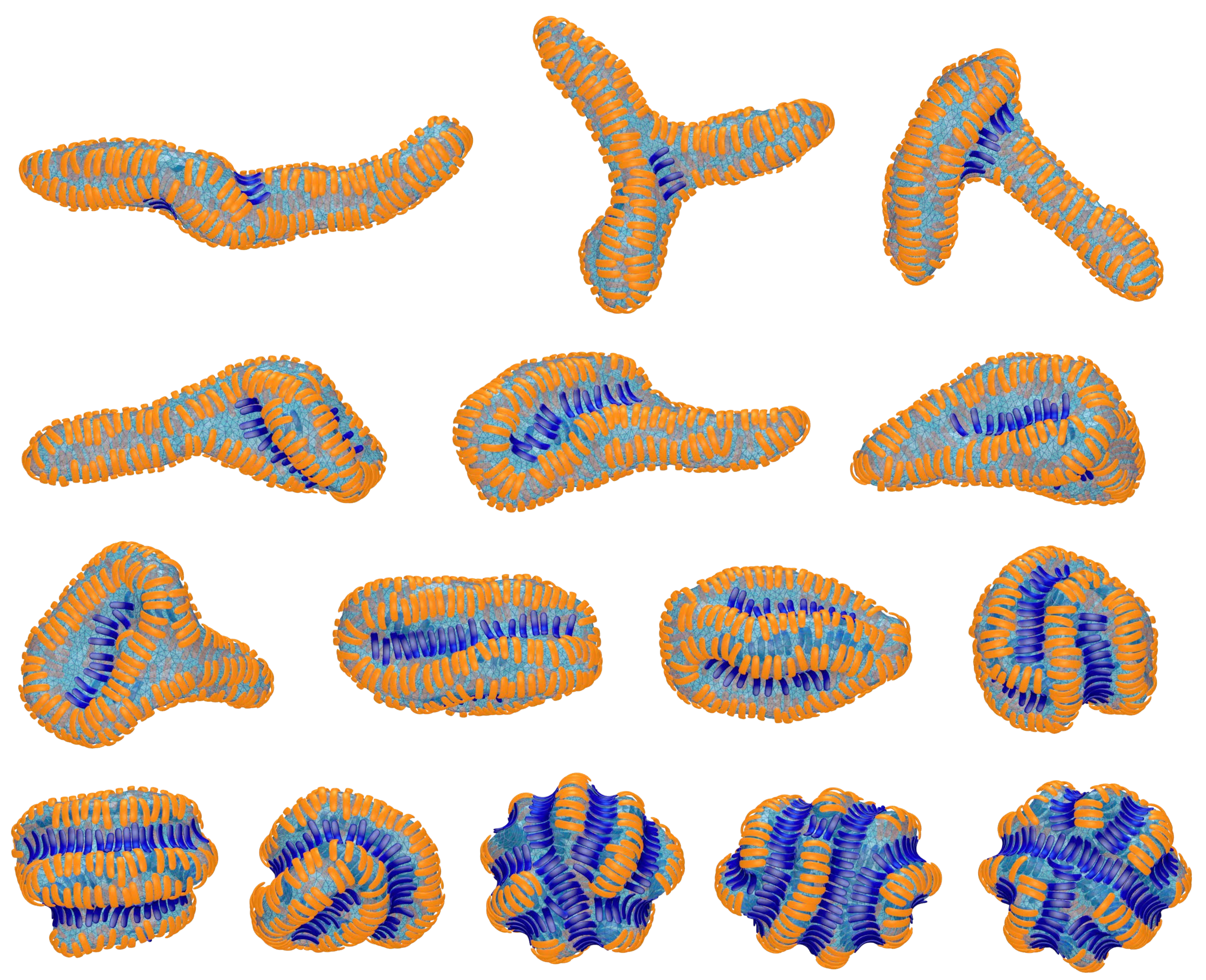}
 \caption{Representative converged morphologies for mixtures of concave and convex particles with arc angle $90\degree$. The morphologies are arranged in ascending order of the membrane coverage of convex, blue particles, which increases from top left to bottom right as $x_{\rm blue} = 0.016, 0.016, 0.016, 0.04, 0.04, 0.04, 0.04, 0.08, 0.08, 0.16, 0.16, 0.16, 0.30, 0.31$, and $0.32$. The membrane coverage of concave, orange particles is $x_{\rm orange} = 0.43, 0.43, 0.42, 0.40, 0.41, 0.39, 0.39, 0.37, 0.37, 0.33, 0.33, 0.33, 0.22, 0.22$, and $0.21$, and the reduced volume of the membrane is $v= 0.50, 0.50, 0.52, 0.54, 0.49, 0.56, 0.54, 0.56, 0.57, 0.69, 0.71, 0.67, 0.73, 0.70$, and $0.71$. The morphologies result from simulations with an initially spherical membrane and the adhesion energy per particle segment $U = 13, 12, 14, 9, 14, 13, 15, 11, 13, 11, 10, 12, 10, 11$, and $12 \, k_B T$.  The overall number of bound and unbound concave particles is $392, 392, 392, 380, 380, 380, 380, 360, 360, 320, 320, 320, 240, 240$, and $240$ in these simulations. The total number of concave and convex particles is $400$ in all simulations.
 }
 \label{figure-4-4}
\end{figure*}

\begin{figure*}[t]
 \centering
\includegraphics[width=\linewidth]{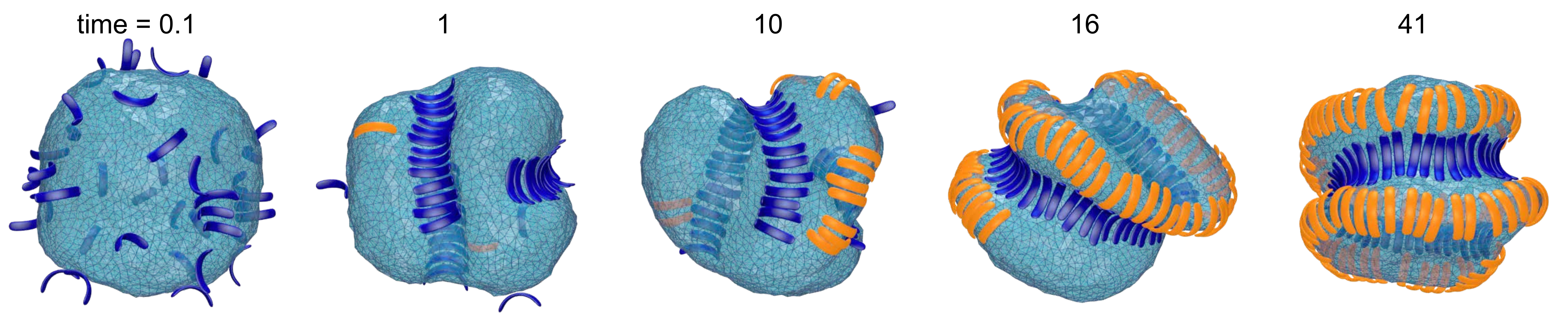}
 \caption{Time sequence of morphologies for a mixture of concave and convex particles with arc angle $120\degree$. The numbers indicate simulation times in units of $10^6$ MC steps per membrane vertex.  In this simulation, the adhesion energy per particle segment is $U = 9\, k_BT$, the total number of concave, orange particles is 359, and total number of convex, blue particles is 41. In the final morphology, 88 out of the 359 concave particles and 40 out of 41 convex particles are bound, which leads to membrane coverages of $x_{\rm orange} = 0.22$ and  $x_{\rm blue} = 0.10$ of the particles. The reduced volume of the membrane in the final morphology is $v = 0.69$.}
 \label{figure-310}
\end{figure*}

\subsection{Simulations}
We have performed Metropolis Monte Carlo (MC) simulations in a cubic box with periodic boundary conditions. The simulations consist of four different types of MC steps: membrane vertex translations, membrane edge flips, particle translations, and particle rotations.\cite{Bonazzi19} Vertex translations enable changes of the membrane shape, while edge flips ensure  membrane fluidity. \cite{Gompper97} In a vertex translation, a randomly selected vertex of the triangulated membrane is translated along a random direction in three-dimensional space by a distance that is randomly chosen from an interval between $0$ and $0.1 a_m$. In a particle translation, a randomly selected particle is translated in random direction by a random distance between $0$ and $a_m$. In a particle rotation, a randomly selected particle is rotated around a rotation axis that passes trough the central point along the particle arc. For particles that consist of 3 or 5 segments, the rotation axis runs through the center of the central segments. For particles composed of 4  segments, the rotation axis runs through the center of the edge that is shared by the two central segments. The rotation axis is oriented in a random direction. The random rotations are implemented using quaternions \cite{Frenkel02,Vesely82} with rotation angles between 0 and a maximum angle of about $2.3\degree$.  Each of these types of MC steps occur with equal probabilities for single membrane vertices, edges, or particles.\cite{Bonazzi19} 

We have run simulations with identical arc angles of either $60\degree$, $90\degree$, or $120\degree$ of the concave and convex particles. The overall number of concave and convex particles in our simulations is $400$, and the initial shape of the membrane is spherical, with all particles unbound. For particles with arc angles of $60\degree$ and $90\degree$, we have run simulations with $8, 20, 40, 80, 160, 240$, and $320$ convex particles out of $400$ particles in total. The adhesion energy per segment is identical for the concave and convex particles in these simulations and has the value $U = 9, 10, 11, 12$, or $13\, k_BT$. In the case of 8 or 20 convex particles, we have also run simulations with  $U = 14$ or $15\, k_BT$. The membrane and particles are enclosed in a cubic simulation box of volume $V_\text{box} \simeq 1.26\cdot 10^5 a_m^3$. To verify convergence, we divide the last $10^7$ MC steps per vertex of a simulation into ten intervals of $10^6$ steps and calculate the reduced volume $v$ of the membrane for each interval. We take a simulation to be converged if the standard deviation of the 10 averages of $v$ for the last 10 intervals of $10^6$ MC steps is smaller than 0.03. The morphologies obtained from converged simulations correspond to metastable or stable states. For the adhesion energies $U \ge 9\, k_BT$ per particle segment considered here, the  total membrane coverage by concave and convex particles after convergence is on average larger than $40$\% for the chosen box size $V_\text{box}$ and total particle number $400$ of our simulations. For total coverages larger than 40\%, the membranes are fully covered by particles. For smaller adhesion energies of $U = 6, 7$, or $8\, k_BT$, the membranes are only partially covered by the particles after convergence, with average total membrane coverages of $2.3$\%, $13$\%, and $29$\%, respectively. For all adhesion energies, intermediate morphologies with partial particle coverage occur in our simulations at early time points prior to convergence, because the particles are initially unbound (see e.g.\ Fig.\ \ref{figure-159}).

%%%
\section{Results}
%%%
\subsection{Particles with arc angles of $60\degree$}

Fig.\ \ref{figure-159} illustrates the segregation and alignment of particles with arc angle $60\degree$ in a simulation with 320 concave, orange and 80 convex, blue particles. All particles are initially unbound in this simulation. After a simulation time of $t = 0.1 \cdot 10^6$ MC steps per membrane vertex, relatively few particles are bound. Some of the bound convex, blue particles are aligned side-to-side in groups of two or three particles, and some of the bound concave, orange particles are aligned in pairs. The alignment of particles of the same type is driven by indirect, membrane-mediated interactions \cite{Weikl18,Idema19,Phillips09,Reynwar07} because the direct particle-particle interactions are purely repulsive in our model. After a simulation time of $10^6$ MC steps per vertex, bound convex, blue particles form continuous lines or grooves along the membrane, and the membrane bulges between these grooves are more sparsely covered by concave, orange particles. The overall coverage of the membrane by particles then increases with time, and the membrane bulges between the grooves of single lines of convex particles are eventually covered by two or three partly irregular lines of concave particles. During the simulation, the reduced volume $v = 6 \sqrt{\pi} V/A^{3/2}\le 1$ of the closed membrane with area $A$ and volume $V$ decreases from values close to 1 to a final value of $v=0.76$. The reduced volume is a measure for the volume-to-area ratio of the closed membrane \cite{Seifert91} and adopts its maximum value of 1 for an ideal sphere. 

The final, converged membrane morphologies depend on the relative coverages of concave and convex particles (see Fig.\ \ref{figure-3-3}). Membranes that are predominantly covered with concave, orange particles as in the first two morphologies of Fig.\ \ref{figure-3-3} adopt a tubular shape. Concave particles with an arc angle of  $60\degree$ induce a transition from a spherical to a tubular membrane shape at a coverage of about 0.4 in the absence of convex particles.\cite{Bonazzi19}  In the first two morphologies, the coverage of concave particles is $x_{\rm orange} = 0.48$ and $0.49$, respectively, while the coverage of convex particles is $x_{\rm blue} = 0.01$. At these small coverages, the convex particles are bound as single particles or pairs in between the concave particles and do not distort the overall tubular shape of the membrane. 
At the coverage $x_{\rm blue} = 0.03$ of the third morphology of Fig.\ \ref{figure-3-3}, the tubular shape of the membrane is distorted by a larger cluster of convex, blue particles.  At the larger coverages $x_{\rm blue}$ of the remaining morphologies of Fig.\ \ref{figure-3-3}, the convex particles form lines along the membrane. If the coverage $x_{\rm orange}$ of the concave particles exceeds the coverage $x_{\rm blue}$ of the convex particles, the membrane morphologies exhibit grooves of single lines of convex particles, and bulges covered by several lines of concave particles in between these grooves. For a coverage $x_{\rm orange}$ of concave particles that is smaller than the coverage $x_{\rm blue}$ of convex particle, grooves are also formed by two parallel lines of convex particles, while bulges in the between the groves can be covered by single lines of concave particles.  The particle lines need to branch or end because the closed membrane vesicle cannot be covered by regular, parallel lines of particles. 

\subsection{Particles with arc angles of $90\degree$ and more}

In the absence of convex particles, concave particles with arc angles of $90\degree$ induce tubules covered by four lines of particles at membrane coverages larger than about  $0.4$. \cite{Bonazzi19} For mixtures of many concave and few convex particles, we observe branched tubule structures as in Fig.\ \ref{figure-209}, with small clusters of convex particles at a three-way junction as branching point. In the simulation of Fig.\ \ref{figure-209}, the number of convex particles is 8, and the total number of bound and unbound concave particles is 392. The bound concave and convex particle have a rather strong tendency to align side-to-side with particles of the same type due to indirect, membrane-mediated interactions.  At the simulation time $t = 1 \cdot 10^6$ MC steps per membrane vertex, bound concave particles form short lines, while convex particles are bound as single particles or in pairs. At time $t = 4 \cdot 10^6$ MC steps per vertex, a line of 5 convex particle is formed. This linear cluster of 5 convex particle remains until time $t = 17 \cdot 10^6$ MC steps per vertex and eventually gains a sixth convex particle at time $t = 22 \cdot 10^6$ MC steps. From time $t = 4 \cdot 10^6$ to  $t = 22 \cdot 10^6$ MC steps per vertex, more and more concave particles bind to the membrane by elongating lines of particles, and these particle lines eventually lead to three tubules protruding from a three-way junction at which the small cluster of convex particles is located. 

The first three of the final, converged morphologies shown in Fig.\ \ref{figure-4-4} result from simulations with same total numbers of 392 concave and 8 convex particles with arc angle $90\degree$  as in the simulation of Fig.\ \ref{figure-209}. In all three morphologies, the 8 convex particles are bound, which leads to the membrane coverage $x_{\rm blue} = 0.016$ of these particles. In the first morphology, the 8 convex particles are bound in a cluster of 4 particles, a cluster of 3 particles, and as a single particle, and induce a distortion or twist in the overall tubular structure induced by the many bound concave particles. In the second and third morphology, the 8 convex particles are bound in two clusters of 4 particles and a single cluster of 8 particles, respectively, which are located at a three-way junction as in Fig.\ \ref{figure-209}. At the larger membrane coverages $x_{\rm blue}= 0.04$ of convex particles in the morphologies 4 to 7 of Fig.\ \ref{figure-4-4}, a tubular protrusion is formed at one end of the closed membrane by bound concave particles, while the remaining membrane is covered by lines of convex and concave particles that induce grooves and bulges. In the remaining morphologies of Fig.\ \ref{figure-4-4}, the membrane is covered by alternating and locally parallel lines of convex and concave particles. Grooves are typically formed by single lines of convex particles, while bulges are covered by either one line or by two parallel lines  of concave particles, depending on the relative coverages of the two particle types. 

For mixtures of concave and convex particles with an arc angle of $120\degree$, we observe a temporal ordering in the binding of the two particle types to an initially spherical membrane (see Fig.\ \ref{figure-310}). At the simulation time $t = 0.1 \cdot 10^6$ MC steps per vertex, only convex particles are bound, and these particles are partially bound with typically one or two of the five segments of which the particles are composed. The partially bound convex particles are not yet aligned and deform the initially spherical membrane only rather slightly. At the simulation time $t =1 \cdot 10^6$ MC steps per vertex, the majority of bound convex particles is fully bound and tightly aligned, which leads to rather deep grooves on the vesicle membrane, and the first concave particles bind to the bulges emerging adjacent to these groves. At time  $t =10 \cdot 10^6$ MC steps per vertex, small linear clusters of concave particles form on the bulges, which eventually grow and coalesce into a single spiral of concave particles that is intertwined with a spiral of convex particles. 

\section{Discussion and Conclusions}

The arc-shaped particles of our model generate membrane curvature by imposing their shape on the membrane upon binding \cite{Bonazzi19}. The arrangements of these particles on the membranes are essentially unaffected by the membrane discretization because the particles are not embedded in the membrane. In other models of membrane shaping \cite{Weikl18,Ramakrishnan18,Simunovic18,Marrink19,Hafner19}, curvature-inducing particles and proteins have been described as nematic objects embedded on the vertices of a triangulated membrane \cite{Ramakrishnan13,Tourdot14}, as curved chains of beads embedded in a two-dimensional sheet of beads that represents the membrane \cite{Noguchi15,Noguchi17,Noguchi17b}, as curved chains of spheres adhered to a triangulated membrane \cite{Helle17}, or as coarse-grained proteins or particles in molecular dynamics simulations. \cite{Reynwar07,Arkhipov09,Simunovic13,Braun14,Simunovic15,Olinger16,Anselmi18,Jarin19,Bhaskara19} Proteins can generate membrane curvature via different mechanisms.  \cite{Shibata09,Kozlov14,McMahon15,Baumgart11,Zimmerberg06} Arc-shaped scaffolding proteins impose curvature on the membrane by binding to the lipid bilayer,\cite{Mim12b,Rao11} transmembrane proteins with a conical or wedged shape induce a curvature on the lipid bilayer that surrounds the proteins,\cite{Phillips09,Aimon14} and hydrophobic protein motifs that are partially inserted into the lipid bilayer can act as wedges to generate membrane curvature.\cite{Campelo08,Boucrot12,Kahraman18} 

 A central parameter for membrane shaping is the induced curvature angle of the particles or proteins.\cite{Bonazzi19,Schweitzer15b} For our arc-shaped particles, the induced angle of the curved membrane segments to which the particles are bound is close to the arc angle of the particles, \cite{Bonazzi19} which varies here from $60\degree$ to $120\degree$. Arc angles of $60\degree$ roughly correspond to the angle enclosed by concave-binding BAR domain proteins such as the Arfaptin BAR domain and the endophilin and amphiphysin N-BAR domains,\cite{Qualmann11,Masuda10} while larger arc angles up to $180\degree$ have been postulated for reticulon scaffolds.\cite{Shibata10,Schweitzer15b} The structural details of the curvature generation by transmembrane proteins such as reticulon and lunapark proteins are not fully known,\cite{Bhaskara19} in contrast to soluble scaffold proteins such as BAR domains. Besides reticulon and lunapark proteins, the generation of the tubular membrane network of the endoplasmic reticulum also requires atlastin proteins, which appear to generate tubular junctions by tethering and fusing tubules \cite{Wang16,Hu09,Orso09}.

The membrane morphologies induced by mixtures of concave and convex particles depend on the relative coverage of these particles, besides the particles' arc angle. For mixtures of few convex and many concave particles with arc angles of $90\degree$, we either find single membrane tubules as in the first morphology of Fig.\ \ref{figure-4-4},
or three tubules connected by a three-way junction as in Fig.\ \ref{figure-209} and in the second and third morphology of Fig.\ \ref{figure-4-4}. These morphologies are formed in simulations with 8 convex and 392 concave particles in total. We have run 7 simulations with these particle numbers for the adhesion energies per segment $U= 9$, $10$, $11$, $12$, $13$, $14$, and $15$, respectively.  In 5 of these 7 simulations, three-way junctions are formed. The few convex particles are bound and clustered in membrane regions of the three-way junction in which the curvature is opposite to the curvature of the tubules that emerge from the junction. The convex particles thus appear to stabilize three-way junctions as suggested for lunapark proteins, which presumably prefer membrane curvature opposite to the tubular curvature.\cite{Chen15} For particles with arc angles of $60\degree$, we do not observe the formation of three-way junctions. One reason may be that the tubes formed by concave particles with an arc angle of $60\degree$ are thicker than tubes induced by concave particles with arc angle  $90\degree$.\cite{Bonazzi19} For the same membrane area, tubes formed by concave particles with arc angle $60\degree$ therefore are shorter, and the finite membrane area in our simulations may impede morphologies with three such thicker tubules emerging from a three-way junction. Another reason is that a few convex particles with arc angle $60\degree$ lead to rather small perturbations of the tubules induced by many concave particles, see the first two morphologies in Fig.\ \ref{figure-3-3}. The convex particles with arc angle $60\degree$ thus are less `disruptive‘ for the tubules, compared to convex particles with arc angle $90\degree$.

For comparable fractions of concave and convex particles, we observe lines of particles of the same type. Lines of convex particles induce membrane grooves, and adjacent, locally parallel lines of concave particles induce bulges next to these grooves. In these lines, the particles are oriented side-to-side. The side-to-side alignment and segregation of the concave and convex particles is driven by indirect, membrane-mediated interactions because the direct particle-particle interactions are purely repulsive in our model.  The segregation patterns of particle lines are reminiscent of the stripe morphologies observed for modulated phases and microphase separation \cite{Seul95}, which arise from a competition of short-range attractive  and long-range repulsive interactions. Here, the segregation into lines of convex and concave particles results from an interplay of particle composition and membrane curvature. The segregation into alternating lines of concave and convex particles appears to be favourable at sufficiently large adhesion energies, because the membrane vesicle can be rather densely covered by the particles of the alternating lines. In addition, there is no line tension between clusters of different particles as driving force for full segregation into two domains of concave and convex particles because of the purely repulsive direct particle-particle interactions in our model. A caveat is that the converged morphologies observed in our simulations correspond to metastable or stable states and, thus, not necessarily to equilibrium states.

In previous work, both side-to-side and tip-to-tip alignment of arc-shaped proteins or particles at membranes has been reported. Attractive membrane-mediated side-to-side pair interactions of arc-shaped particles have been obtained from energy minimization \cite{Schweitzer15a}. Side-to-side alignment has also been observed in simulations with arc-shaped inclusions in membranes \cite{Noguchi15,Noguchi17,Noguchi17b}. In molecular dynamics (MD) simulations with a coarse-grained molecular model of N-BAR domains proteins on DLPC lipid vesicles, in contrast, a tip-to-tip alignment of proteins has been observed \cite{Simunovic13,Simunovic17}, which may be affected by the direct, coarse-grained protein-protein interactions of the model. A tip-to-tip alignment has also been reported for MD simulations with a coarse-grained model of I-BAR domains \cite{Jarin19} and for coarse-grained MD simulations of arc-shaped nanoaparticles on lipid vesicles at large adhesion energies of the nanoparticles \cite{Olinger16}. At these large adhesion energies, the nanoparticles are partially wrapped by the membrane, which leads to saddle-like membrane curvature around nanoparticles that may cause side-to-side repulsion. At  smaller adhesion energies, the arc-shaped nanoparticles induce membrane curvature only along their arcs and align side-to-side, similar to our arc-shaped particles.
In simulations with mixtures of arc-shaped and conical inclusions in membranes, the tubulation caused by the arc-shaped particles has been found to be accelerated if the conical inclusions induce curvature of the same sign, and suppressed if the conical inclusions induce curvature of opposite sign \cite{Noguchi17b}. For mixtures of arc-shaped inclusions with opposite curvatures, adjacent lines of the different particles have also been observed at overall relatively low densities of the particles \cite{Noguchi17}. 

The morphologies in our simulations result from an interplay of the bending energy of the membrane and the overall adhesion free energy of the particles. In these simulations, the membranes are tensionless because the volume enclosed by the membrane is not constrained, in order to allow for a wide range of morphologies with different volume-to-area ratios.  In general, the bending energy dominates over the membrane tension $\sigma$ on length scales smaller than the characteristic length $\sqrt{\kappa/\sigma}$, which adopts values between 100 and 400 nm for typical tensions $\sigma$ of a few $\mu\text{N}/\text{m}$ \cite{Simson98,Popescu06,Betz09} and typical bending rigidities $\kappa$ between $10$ and $40$ $k_B T$.\cite{Nagle13,Dimova14}  Our results thus hold on length scales smaller than this characteristic length. In contrast, the overall membrane morphology on length scales larger than $\sqrt{\kappa/\sigma}$ depends on the membrane tension \cite{Lipowsky13,Shi15,Simunovic15}.

\section*{Conflicts of interest}
There are no conflicts to declare.

\section*{Acknowledgements}
Financial support from the Deutsche Forschungsgemeinschaft (DFG) {\em via} the International Research Training Group 1524 ``Self-Assembled Soft Matter Nano-Structures at Interfaces" is gratefully acknowledged.

%%%END OF MAIN TEXT%%%


\balance


%%%REFERENCES%%%
%\bibliography{membranes} %You need to replace "rsc" on this line with the name of your .bib file
%\bibliographystyle{rsc} %the RSC's .bst file

\providecommand*{\mcitethebibliography}{\thebibliography}
\csname @ifundefined\endcsname{endmcitethebibliography}
{\let\endmcitethebibliography\endthebibliography}{}

\end{document}